\documentclass[11pt]{article}
\usepackage{blois}
\usepackage{amsmath,amssymb,mathrsfs}

\def\be{\begin{equation}}
\def\ee{\end{equation}}
\def\bea{\begin{eqnarray}}
\def\eea{\end{eqnarray}}

\newcommand{\Photo}{}

\begin{document}
\vspace*{4cm}
\title{ULTRALIGHT PARTICLE DARK MATTER}

\author{ A. RINGWALD}

\address{Deutsches Elektronen-Synchrotron DESY, Notkestrasse 85,\\
D-22607 Hamburg, Germany}

\maketitle\abstracts{
We review the physics case for very weakly coupled ultralight particles beyond the Standard Model, 
in particular for axions and axion-like particles (ALPs): (i) the axionic solution of the strong CP problem 
and its embedding in well motivated extensions of the Standard Model;
(ii) the possibility that the cold dark matter in the Universe is comprised of axions and ALPs; 
(iii) the ALP explanation of the anomalous transparency of the Universe for TeV photons; and 
(iv) the axion or ALP explanation of the anomalous energy loss of white dwarfs.
Moreover, we present an overview of ongoing and near-future laboratory 
experiments searching for axions and ALPs: haloscopes, helioscopes, and light-shining-through-a-wall
experiments.
}

\section{Introduction}

There are plenty of particle candidates for the cold dark matter in the Universe, spanning a 
huge parameter range in masses and strengths of couplings to the particles of the Standard Model. 
Among them, two particular classes stand out because of their convincing physics case and the variety 
of possible experimental and observational opportunities to search for them:  Weakly Interacting Massive Particles (WIMPs), such as 
the neutralinos in supersymmetric extensions of the Standard Model, and 
very Weakly Interacting Slim (in the sense of ultralight) Particles (WISPs), such as axions in 
Peccei-Quinn extensions of the Standard Model.
In this contribution to the conference, we will concentrate on WISPs. In particular, we will review the 
physics case for axions and axion-like particles and present ongoing and near-future laboratory experiments hunting for them. For more details and a review of other WISP candidates, see 
Refs. \cite{Jaeckel:2010ni,Ringwald:2012hr,Jaeckel:2013ija}.

\section{Physics Case for Axions and Axion-like Particles}

\subsection{The Axionic Solution of the Strong  CP Problem}

The axion is predicted in an elegant solution of the so-called strong CP problem. It is the prime example of a WISP. 

Shortly after the emergence of Quantum Chromo-Dynamics (QCD) as the fundamental theory of strong interactions it was realized that 
its fundamental parameters are not only the strong fine structure constant $\alpha_s$ and the quark masses 
$m_u$, $m_d$, \ldots, but also the angular parameter 
$\overline{\theta} = \theta + {\rm arg\ det}\ {\mathcal M}_q$, 
a sum of the $ {\rm arg\ det}$ of the quark mass matrix 
${\mathcal M}_q$   and  the QCD vacuum angle $\theta$, appearing in the theta term 
in the QCD Lagrangian, 
\begin{equation}
{\mathcal L} \supset  -  \frac{\alpha_s}{8\pi} \,\theta\, G_{\mu\nu}^a {\tilde G}^{a,\mu\nu} ,
\label{qcd}
\end{equation} 
where $G_{\mu\nu}^a$ is the gluon field strength and  ${\tilde G}^{a,\mu\nu}$ its dual.   
This term 
$\propto G_{\mu\nu}^a {\tilde G}^{a,\mu\nu} \propto \mathbf{E}^a\cdot \mathbf{B}^a$,  violates T and P, and thus CP. 

Experimentally, the most sensitive probe of T and P violation in flavor conserving interactions is the electric dipole moment (EDM) of the neutron, which turns out to be tiny \cite{Baker:2006ts},
$\left| d_n\right| < 2.9\times 10^{-26}\ e\,{\rm cm}$,
implying an exponentially small value of the theta parameter,
$\left| \overline{\theta}\right| \lesssim 10^{-9}$,
obtained from comparing the QCD prediction,
\begin{equation}
d_n (\overline{\theta})\sim   \frac{e  m_u m_d\overline{\theta}}{(m_u + m_d)m_n^2}
\sim 6\times 10^{-17}\ \overline{\theta}\ e\,{\rm cm},
\end{equation}
with the experimental value. 
This tremendous fine-tuning of the angular theta parameter -- the latter being expected to 
be naturally of order unity -- constitutes the strong CP problem. 

The Peccei-Quinn (PQ) solution \cite{Peccei:1977hh} of the strong CP problem is based on the observation that the vacuum energy in QCD, 
\begin{equation}
V(\overline{\theta})=\frac{ m_\pi^2 f_\pi^2}{2} 
\frac{m_u m_d}{(m_u+m_d)^2} 
\overline{\theta}^2 + {\mathcal O}(\overline{\theta}^4)
, \label{effpot}
\end{equation}
where $m_\pi$ and $f_\pi$ are the mass and the decay constant of the pion, 
respectively, has a localized minimum at vanishing theta parameter. Therefore, 
if theta were a dynamical field, its vacuum expectation value (vev) would be zero.

The idea is then to introduce a field $a(x)$, respecting a non-linearly realized $U(1)_{\rm PQ}$ symmetry, 
i.e. a shift symmetry, $a(x)\to a(x) + {\rm const.}$, broken only by the coupling to $G\tilde{G}$,
\begin{equation}
{\mathcal L} \supset  -  \frac{\alpha_s}{8\pi} \,
\left(\overline{\theta}+\frac{a}{f_a}\right)\, G_{\mu\nu}^a {\tilde G}^{a,\mu\nu} ,
\label{aqcd}
\end{equation} 
where the Peccei-Quinn (PQ) scale $f_a$ is a parameter with dimension of mass. The pseudo-scalar field $a(x)$ 
thus plays effectively the role of a dynamical theta parameter. 
In fact, exploiting the shift symmetry one can eliminate $\overline{\theta}$ in Eq. (\ref{aqcd}).
QCD dynamics, cf. Eq. (\ref{effpot}), leads then to a vanishing  vev for the shifted field, i.e. T, P, and CP are
conserved. 

Inevitably, this mechanism predicts the existence of a particle excitation corresponding to the 
quantum field fluctuations 
around the vev:  the  ``axion" \cite{Weinberg:1977ma,Wilczek:1977pj}. 
The axion is a pseudo-Nambu-Goldstone boson rather than a proper, i.e. massless, Nambu-Goldstone boson: 
because of mixing with the neutral pion, described by the chiral QCD Lagrangian, the shift symmetry is
broken and the axion acquires a mass, which can be read off from the quadratic part in the potential (\ref{effpot}),   
\begin{equation}
m_a = \frac{m_\pi f_\pi}{f_a} \frac{\sqrt{m_u m_d}}{m_u+m_d}\simeq { 0.6\,  {\rm meV}}
         \times
         \left(
         \frac{10^{10}\, {\rm GeV}}{f_a}\right).
\label{axion_mass}
\end{equation}
 
\begin{figure}
\centerline{\includegraphics[width=0.6\linewidth]{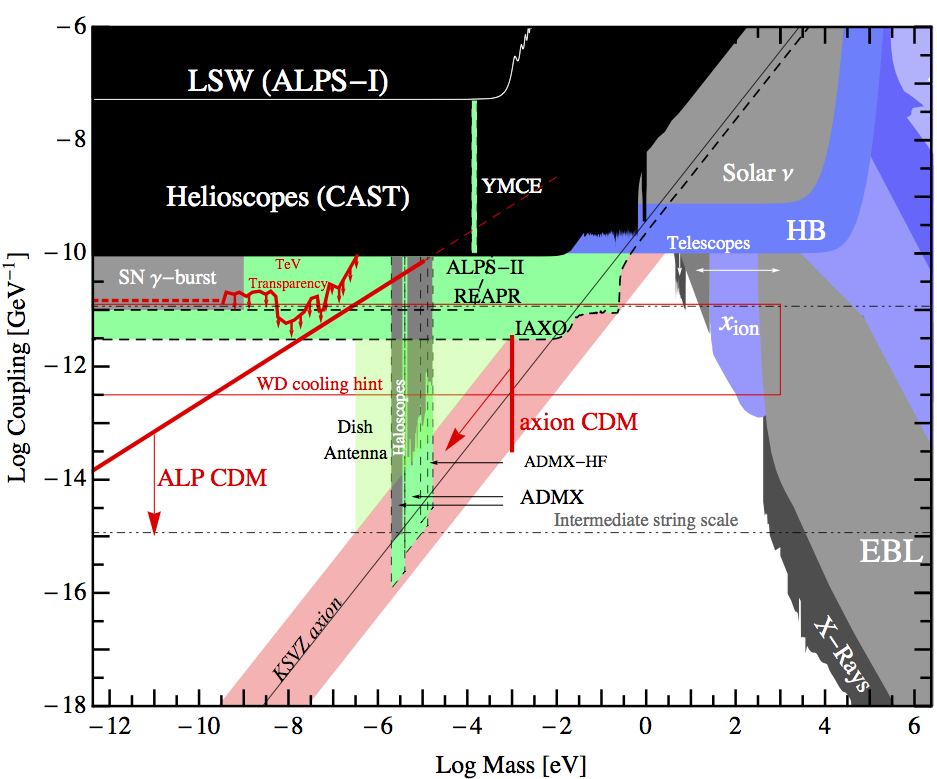}}
\caption{Exclusion regions in axion and ALP coupling to the photon vs. mass as described in the text.}
\label{fig:alpplot}
\end{figure}

The interactions of the axion with the Standard Model particles are suppressed by inverse powers of the 
PQ scale, as follows from the shift symmetry and behooves for a Nambu-Goldstone boson. From its mixing with the pion it inherits a universal coupling to hadrons and to photons, 
\begin{eqnarray}
{\cal L} \supset - \frac{1}{4}\, g_{a\gamma}\, a\, F_{\mu \nu}\tilde{F}^{\mu \nu} =
g_{a\gamma}\, a\, {\mathbf E}\cdot {\mathbf B},
\end{eqnarray}
with 
\begin{eqnarray}
        { g_{a\gamma}} = \frac{\alpha}{2\pi f_a}
\left( C_{a\gamma} - {\frac{2}{3}\,\frac{m_u+4 m_d}{m_u+m_d}     }\right)
\sim \left( C_{a\gamma} -  1.92\right)\times 
10^{-13}\ {\rm GeV}^{-1}   \left(
         \frac{10^{10}\, {\rm GeV}}{f_a}\right)       ,
         \label{axionphotoncoupling}
\end{eqnarray}
where $C_{a\gamma}$ is a model dependent dimensionless factor. This model dependence is illustrated
by the width of the red band in Fig. \ref{fig:alpplot} (taken from Ref. \cite{Baker:2013zta}), which displays  
the expected axion coupling to the photon as a function of its mass $g_{a\gamma}\propto 1/f_a\propto m_a$.  
 
The phenomenology of the axion is therefore largely determined by the value of the PQ scale. 
Currently, the most stringent lower bounds on it arise from astrophysical considerations and 
are summarized in Fig. \ref{fig:axionplot} which has been adapted and updated from Ref. \cite{Hewett:2012ns}.
In particular, the observed duration of the neutrino signal from
supernova SN 1987A pose the most stringent model independent limit,  
$
f_a > 4\times 10^8\,{\rm GeV} \Rightarrow 
m_a < 16\,{\rm meV}$.
Thus, the axion is a prime paradigm of a WISP: it is indeed ultralight and 
very weakly coupled to the Standard Model.

\subsection{Ultraviolet Completions of the Standard Model Predicting Axions and Axion-Like Particles}

\begin{figure}
\centerline{\includegraphics[width=0.6\linewidth]{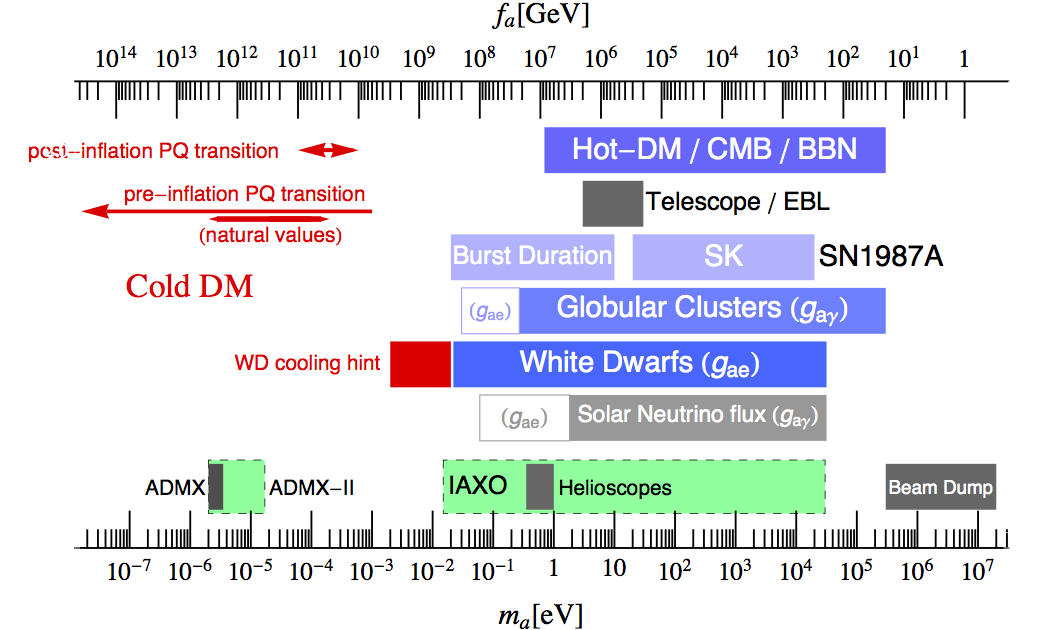}}
\caption{Exclusion regions in Peccei-Quinn scale $f_a$ and axion mass $m_a$ as described in the text.}
\label{fig:axionplot}
\end{figure}

In purely field theoretic high-energy extensions of the Standard Model in four dimensions, 
the axion can be obtained as a Nambu-Goldstone boson from the breaking of a 
global $U(1)_{\rm PQ}$ symmetry. One introduces a complex 
Standard Model singlet scalar field $\sigma$ whose vev 
$\langle \sigma\rangle =v_{\rm PQ}/\sqrt{2}$
breaks $U(1)_{\rm PQ}$ and whose phase, cf. $
\sigma (x) =\frac{1}{\sqrt{2}}\big[v_{\rm PQ}+\rho (x)\big]e^{ia(x)/v_{\rm PQ}}$, is the axion field.
The PQ scale $f_a$ is then essentially the $U(1)_{\rm PQ}$
symmetry breaking scale, 
$f_a = v_{\rm PQ}/C_{ag}$,
where $C_{ag}$ is a model dependent factor of order unity which depends on the 
$U(1)_{\rm PQ}$ charge assignments of the fermions in the theory which have to be 
such that the mixed $U(1)_{\rm PQ}\times SU(3)_C\times SU(3)_C$ axial anomaly is non-vanishing. 
Very well known field theoretic PQ extensions of the Standard Model are the so-called KSVZ  \cite{Kim:1979if,Shifman:1979if}  and DFSZ  \cite{Dine:1981rt,Zhitnitsky:1980tq} models.
The PQ breaking scale in these models is however still ad-hoc. Interestingly, there are a number
of well-motivated models which extend the Standard Model even more, thereby tying the PQ scale with other supposedly fundamental scales. Particularly intriguing are models in which additionally right-handed 
singlet Majorana neutrinos are introduced which carry also PQ charges and get their mass by PQ symmetry breaking, such that the PQ scale can in fact be identified with the seesaw scale. The masses of the active 
neutrinos, $m_\nu\propto v^2/v_{\rm PQ}$, where $v=246$\,GeV is the electroweak breaking scale, are then naturally of order the observed mass splittings, $\sim 0.01$\,eV, for an intermediate scale 
$v_{\rm PQ}\sim 10^{10\div 12}$\,GeV \cite{Langacker:1986rj}. 
These models, which can be embedded in certain 
Grand Unified Theories (GUTs) \cite{Mohapatra:1982tc,Dias:2004hy,Altarelli:2013aqa},  therefore indeed prefer a PQ scale in the allowed parameter range, cf. Fig. \ref{fig:axionplot}. 

Still, in these field theoretic PQ extensions one has to introduce the complex scalar singlets and 
the global $U(1)_{\rm PQ}$ by hand. This is different in low energy effective field theories emerging from
 string theory. 
In fact, they contain automatically candidates for the axion, often even an ``axiverse" containing
additional axion-like particles (ALPs), the latter being characterized by the fact that 
they do not couple to $G \tilde{G}$. 
One class, the so-called closed string axions and ALPs, arise as Kaluza-Klein zero modes 
of antisymmetric tensor fields living in ten dimensions and inherit their shift symmetry from 
gauge invariance in ten dimensions. Their number depends on the topology
of the compactified six-dimensional space. Their predicted PQ scales $f_{a/{\rm ALP}}$ are typically of order of the
string scale~\cite{Witten:1984dg,Arvanitaki:2009fg,Acharya:2010zx}. The latter is close to the Planck scale, $\sim 10^{18}$\,GeV, in compactifications of the heterotic string, or the intermediate scale, $\sim 10^{10\div 12}$\,GeV, in LARGE volume compactifications of type II string theory realising brane world 
scenarios~\cite{Cicoli:2012sz}. 
Further axions and ALPs appear automatically as pseudo Nambu-Goldstone bosons from the breaking of accidental $U(1)_{\rm PQ}$ symmetries that appear as low energy remnants of exact discrete symmetries occuring in compactifications of the heterotic string~\cite{Lazarides:1985bj,Choi:2009jt,Dias:inprep}.  
In this case, it is possible to disconnect 
the PQ scales $f_{a/{\rm ALP}}$ from the string scale
and chose them, e.g., in the phenomenologically interesting intermediate scale region.

\subsection{Axions and ALPs as Cold Dark Matter}

A few years after the proposal of the Peccei-Quinn solution of the strong CP problem 
it was realized that, for $f_a \gtrsim 10^{10}$\,GeV, the axion is not as invisible as previously 
thought. In fact, for such large values of the PQ scale, it is produced mainly 
non-thermally in the early Universe and may comprise the dominant part of
cold dark matter (for a review, see Ref.  \cite{Sikivie:2006ni}). 

The most generic non-thermal production in the early Universe proceeds via the so-called 
vacuum-realignment mechanism \cite{Preskill:1982cy,Abbott:1982af,Dine:1982ah}.
When the temperature in the early universe drops below 
the PQ scale, the spatially homogeneous mode of the axion field is frozen at a random initial value, 
$\theta_i = a(t_i)/f_a$, in each different causally connected region.  Later, at the epoch of 
chiral symmetry breaking, the axion potential (\ref{effpot}) rises and the axion acquires its
mass $m_a$. Then, the axion field relaxes to its CP conserving minimum, around which it oscillates. 
These oscillations represent a population of non-relativistic axions.  

If the reheating temperature after inflation is above the PQ scale, the initial misalignment angles 
$\theta_i = a(t_i)/f_a$ are randomly distributed in different causally connected parts of the Universe,
leading to a prediction of 
\begin{equation}
\Omega_a h^2\approx  0.3 \times \left( \frac{f_a}{10^{12}\ \rm GeV} \right)^{7/6} 
\label{eq:omegaqcdaxionav}
\end{equation}
for the cosmic mass fraction in axions. Therefore, in this case, 
$m_a \sim 10^{-5}$~eV axions,
corresponding to a PQ scale of order $f_a\sim  10^{11}$\,GeV, 
can provide dark matter, but smaller masses (higher PQ scales) are excluded (in the absence of a dilution mechanism by late decays of particles beyond the Standard Model). 
Apart from the vacuum-realignment mechanism, the decay of cosmic strings and  domain walls may provide 
additional sources for non-thermally produced axions~\cite{Sikivie:2006ni}. 
These extend the region where axions may provide the whole of cold dark matter 
in the Universe to  $10^{10}\ {\rm GeV}\lesssim f_a\lesssim 10^{11}$~GeV~\cite{Wantz:2009it,Hiramatsu:2010yu,Hiramatsu:2012gg}, cf. the red region 
labelled ``post-inflation PQ transition" in Fig. \ref{fig:axionplot}.

On the other hand, if the reheating temperature after inflation is below the PQ scale, then 
the axion field is homogenized by inflation, the inital misalignment angle thus being unique in all the observable Universe. Furthermore,  topological defects are diluted away by inflation. Correspondingly,
only the vacuum-realignment mechanism will contribute to the expected cosmic mass fraction in 
axions, which is obtained as 
\begin{equation}
\Omega_a h^2\approx  0.71 \times \left( \frac{f_a}{10^{12}\ \rm GeV} \right)^{7/6} \left( \frac{\theta_i}{\pi} \right)^2,
\label{eq:omegaqcdaxion}
\end{equation}    
if there is no dilution by, e.g., late decays of particles beyond the Standard Model. 
Therefore, the axion can be the dominant part 
of cold dark matter if its decay constant exceeds $f_a \gtrsim 10^{11}$~GeV, corresponding to a mass 
$m_a < 10^{-5}$~eV. 
An axion with a GUT scale decay constant, $f_a\sim 10^{16}$~GeV, on the other hand, would
overclose the universe, unless the initial misaligment angle is very small, $\theta_i\sim 10^{-3}$ 
(``anthropic axion window"~\cite{Tegmark:2005dy}), or there is a late dilution.

Similar to axions, also ALPs would be produced 
in the early Universe via the vacuum-realignment mechanism, leading to an expected cosmic mass fraction~\cite{Arias:2012mb} 
\begin{equation}
\Omega_{a_i} h^2 \approx 0.16 \times
\left( \frac{m_{\rm ALP}}{\rm eV} \right)^{1/2} \left( \frac{f_{\rm ALP}}{10^{11}\ \rm GeV} \right)^{2}  \left( \frac{\theta_i}{\pi} \right)^2,
\label{eq:omegaalp}
\end{equation}
if the reheating temperature after inflation is below $f_{\rm ALP}$. This enlarges the 
search space for ultralight particle dark matter considerably, cf. Fig. \ref{fig:alpplot}.

%We conclude that the natural range for axion/ALP CDM, the  “cosmic axion window”,  

\subsection{ALP Explanation of Anomalous Transpareny of the Universe for TeV Photons}

Intriguingly, the existence of at least one ALP could help to solve a puzzle in gamma ray astronomy,
namely the non-observation of absorption features in very high energy (VHE) photon spectra from distant active galactic nuclei (AGN) which are expected due to $e^+ e^-$ pair production off the extragalactic background light \cite{Horns:2012fx}.
This may be explained by photon $\leftrightarrow$ ALP oscillations:
the conversion of gamma rays into ALPs in the
magnetic fields around the AGNs or in the intergalactic medium, followed by their unimpeded
travel towards our galaxy and the consequent reconversion into photons in the galactic/intergalactic magnetic
fields~\cite{DeAngelis:2007dy,Simet:2007sa,SanchezConde:2009wu}, requiring a very light ALP, which couples to two photons with strength \cite{Meyer:2013pny}
$g_{{\rm ALP}\,\gamma} \gtrsim 10^{-12} \ {\rm GeV}^{-1} 
\Rightarrow f_{\rm ALP}/C_{{\rm ALP}\,\gamma} \lesssim 10^9\ {\rm GeV}$, for
$m_{\rm ALP} \lesssim 10^{-7}\  {\rm eV}$.
Note, that this particle must be an ALP, rather than the axion. In fact, an axion with $m_a \lesssim 10^{-7}$\,eV would have a photon coupling
$g_{a\gamma} \lesssim 10^{-16}$\,GeV$^{-1}$.

\subsection{Axion or ALP Explanation of Anomalous Energy Loss of White Dwarfs}

Bremsstrahlung of axions or ALPs, if they couple directly to electrons, may also explain another puzzle in astrophysics. 
In fact,  the white dwarf (WD) luminosity function seems to require a new energy loss 
channel that can be interpreted in terms of axion or ALP losses corresponding to a PQ scale 
$f_{a/{\rm ALP}}/C_{a/{\rm ALP}\,e}\simeq
(0.7\div 2.6)\times 10^9\ {\rm GeV}$,
where $C_{a/{\rm ALP}\,e}$ is a model-dependent dimensionless constant~\cite{Isern:2008nt}.
Moreover, the observations of the period decrease of the pulsating WDs 
G117-B15A and R548 imply additional cooling that can be
interpreted in terms of similar axion/ALP losses \cite{Isern:2010wz,Corsico:2012sh}. 
The corresponding region in axion parameter space is shown as a red box labelled as ``WD cooling hint" in Fig.~\ref{fig:axionplot}. 
The red parameter region labelled ``WD cooling hint" in Fig.~\ref{fig:alpplot} indicates 
the extrapolation to the  two-photon coupling, if one allows for a reasonable uncertainty 
in the ratio $C_{a/{\rm ALP}\,\gamma}/C_{a/{\rm ALP}\,e}$.

%Hints of relativistic ALP background (dark radiation) in CMB, generated by modulus decay; ALP decay to %photons in magnetic fields of galaxy clusters, e.g. COMA, may explain observed soft X-ray excesses        %[Marsh,Conlon 13]

\section{Laboratory Searches for Axions and ALPs}

Fortunately, there appear to be a number of opportunities for laboratory searches for axions and ALPs if 
their  
photon couplings are indeed in the range $10^{-16}$\,GeV$^{-1}\lesssim g_{a/{\rm ALP}\gamma} \lesssim 10^{-11}$\,GeV$^{-1}$, corresponding to intermediate scale PQ scales, 
$f_{a/{\rm ALP}}\sim 10^{9\div 12}$\,GeV, which we will review in the next few 
subsections. A common feature of these experiments is that they exploit 
the conversion of axions and ALPs into photons (or the reverse processes) in strong, laboratory made magnetic fields.

\subsection{Haloscope Searches}

These searches aim for the direct detection of dark matter axions or ALPs in the laboratory. 

The currently most sensitive haloscopes consist of a high-Q microwave cavity placed in  
a superconducting solenoidal magnet~\cite{Sikivie:1983ip}, trying to detect the electromagnetic power arising from
the conversion of dark matter axions or ALPs into real photons, with frequency 
$
\nu=m_{a/{\rm ALP}}/(2\pi )=0.24\ {\rm GHz}\times (m_{a/{\rm ALP}}/\mu{\rm eV})$.
The best sensitivity is reached on resonance, the power output then being proportional to the quality factor of the cavity. A number of experiments of this type have already been done and the limits obtained 
to date, scanning the mass range by tuning the cavities, are shown as a grey region labelled ``Haloscopes" in Fig. \ref{fig:alpplot}. 
The Axion Dark Matter eXperiment (ADMX) has indeed reached the sensitivity to probe 
proper axion dark matter~\cite{Asztalos:2009yp} (see also Fig. \ref{fig:axionplot}).
The ongoing experiments ADMX-II and ADMX-High Frequency (HF) aim to explore the correspondingly labelled green regions  in the same figures and will thus probe a significant part of the natural parameter region in scenarios where the PQ transition happens before inflation. Further haloscope opportunities
in complementary mass ranges 
may arise from recycling available microwave cavities and magnets at accelerator laboratories~\cite{Baker:2011na}. One example is the WISPDMX proposal in Hamburg, which plans to use a high-Q HERA accelerator cavity and a superconducting solenoid from the H1 experiment \cite{Horns:2013ira}.
 
Complementary to these narrow-band microwave cavity based searches, dish antennas have also been proposed as haloscopes \cite{Horns:2012jf}.  In fact, monochromatic photons generated by the conversion of dark matter axions/ALPs will be emitted perpendicular to the surface of a spherical dish antenna, if it is placed in a background magnetic field. The photons can be focused in the centre of the dish onto a broad-band detector. 
The projected sensitvity of such an haloscope is shown in Fig. \ref{fig:alpplot} as a green region
labelled as ``Dish Antenna". 

Another possibility to search directly for axion/ALP cold dark matter  may be based on the fact that the oscillating galactic dark matter axion/ALP fields will induce oscillating nuclear electric dipole moments in the
laboratory.  Conceivably, these could be detected by precision magnetometry, most sensitive in the sub-MHz band, corresponding to  $m_{a/{\rm ALP}}\lesssim $~neV,
which means trans-GUT scale, $f_{a}\gtrsim 10^{16}$~GeV, for the axion~\cite{Graham:2013gfa,Budker:2013hfa}.
 
\subsection{Helioscope Searches}

These searches aim to detect 
solar axions and ALPs by their conversion into photons inside of a strong magnet pointing towards the Sun~\cite{Sikivie:1983ip}. In fact, up to 10\% of the solar photon luminosity could be radiated in 
axions of ALPs without
being in conflict with helioseismological and solar neutrino data.
The currently best sensitivity has been obtained
by the CERN Axion Solar Telescope (CAST), employing an LHC dipole test magnet. As can be seen in Fig.~\ref{fig:alpplot}, 
the CAST limits (labelled ``Helioscopes (CAST)") have already surpassed, at low masses, 
the bounds from Horizontal Branch (HB) stars~\cite{Andriamonje:2007ew}. 
At larger masses, CAST has started to probe the predictions for 
axions~\cite{Arik:2011rx,Arik:2013nya}.  
A proposed next-generation axion helioscope~\cite{Irastorza:2011gs}, dubbed the International Axion Observatory (IAXO),
envisions a dedicated superconducting toroidal magnet with much bigger aperture than CAST,
a detection system consisting of large X-ray telescopes coupled to
ultra-low background X-ray detectors, and a large, robust tracking system \cite{Vogel:2013bta}. 
IAXO would enable
the search for solar axions and ALPs in a broad mass range, as shown by the green region 
labelled ``IAXO" in Fig.~\ref{fig:alpplot}, allowing to probe a part of axion parameter space, 
corresponding to $f_a\sim 10^{9\div 10}$~GeV, which will not be covered by the haloscope searches.
The projected sensitivity of IAXO nicely overlaps with the ALP parameter region required to
explain the puzzles from astrophysics, such as the anomalous transparency of the universe for VHE 
photons, see Fig.~\ref{fig:alpplot}.

\subsection{Light-Shining-through-a-Wall (LSW) Searches}  

These searches aim for the production and detection of axions or ALPs in the laboratory. 
In these experiments, laser photons are sent along a strong magnetic field, allowing for their conversion into axions or ALPs, which may then reconvert  
in the strong magnetic field behind a blocking wall into photons, susceptible to detection~\cite{Anselm:1986gz,VanBibber:1987rq,Redondo:2010dp}. 
The currently best sensitivity 
has been reached by the Any Light Particle Search (ALPS-I) 
experiment at DESY~\cite{Ehret:2010mh}, which used 
a superconducting HERA dipole magnet, with the first half of the magnet,
the production region, being enclosed in an optical resonator 
and the second half being used as the regeneration region. The end of the
latter was connected to a light-tight box in which signal photons were redirected to
a CCD camera. The bound established by ALPS-I for ALPs is labelled ``LSW (ALPS-I)"  
in Fig.~\ref{fig:alpplot}. 
Recently, the first stages of the successor experiment ALPS-II have been approved by the 
DESY directorate \cite{Bahre:2013ywa,Dobrich:2013mja}. ALPS-II proposes to use 10+10 straightened 
HERA magnets \cite{Ringwald:2003nsa}, a high-power laser
system, a superconducting low-background detector and the pioneering realization of an optical
regeneration cavity~\cite{Hoogeveen:1990vq}. The Resonantly Enhanced Axion-Photon Regeneration Experiment (REAPR) at Fermilab is a quite similar proposal which differs from ALPS-II in the
scheme to lock the optical 
cavities~\cite{Mueller:2009wt}.
This generation of LSW experiments aims 
at surpassing the current helioscope bounds from CAST and tackle some of the ALP 
parameter space favored by astrophysical observations, cf. the light-green region in 
Fig.~\ref{fig:alpplot} labelled by ``ALPS-II, REAPR". 

%LSW experiments can also be performed in other spectral ranges, notably in the 
%microwave~\cite{Hoogeveen:1992nq,Jaeckel:2007ch,Caspers:2009cj} 
%and in the X-ray ranges~\cite{Rabadan:2005dm,Dias:2009ph}. 

\section{Summary and Conclusions}

There is a strong physics case for very weakly coupled ultralight particles beyond the Standard Model. 
The elegant solution of the strong CP problem proposed by Peccei and Quinn gives a particularly strong motivation for the axion. In many 
theoretically appealing ultraviolet completions of the Standard Model, in particular in 
completions arising from strings, there occur axions and axion-like particles 
(ALPs) automatically. Moreover, axions and ALPs are natural cold dark matter
candidates. They can explain the anomalous transparency of the Universe for VHE gamma rays
and the anomalous cooling of white dwarfs. 
Fortunately, a significant portion of axion and ALP parameter space will be tackeled in this 
decade by a number of experiments of haloscope, helioscope, and light-shining-through-a-wall type. 
Stay tuned!

\end{document}